\documentstyle[preprint,aps]{revtex}

\begin{document}
\thispagestyle{empty}
\begin{titlepage}
\draft
\preprint{OHSTPY-HEP-T-94-020}
\title{Rephase-Invariant CP-Violating Observables and Mixings \\
in the $B^{0}$-, $D^{0}$- and $K^{0}$- Systems
\footnote{work supported in part by US Department of Energy grant
DOE/ER/01545-605}   }
\author{W.F. Palmer and   Y.L. Wu}
\address{ Department of Physics, \ Ohio State  University \\ Columbus,
 Ohio 43210,  \ U.S.A.}
\date{Dec. 1994}
\maketitle

\begin{abstract}
 We present a general model-independent and rephase-invariant formalism that
cleanly relates observables to the fundamental parameters and explicitly
separates different types of CP violation in the $B^0$-, $D^0$- and $K^0$-
systems. We emphasize its importance when interpreting experimental measurement
of CP violation, the unitarity triangle, and probes of  new physics.
\end{abstract}

\end{titlepage}

\narrowtext

CP violation in neutral meson decays arises from CP violating phases in the
mixing matrix (indirect CP violation) or in the weak decay amplitudes (direct
CP violation).
In the Cabbibo, Kobayashi and Maskawa (CKM) \cite{CKM} model of
CP violation, both direct and indirect CP violation occur. They can be
measured by studying time evolution of neutral meson
decays \cite{BS,WOLF,PZ,DR,BKUS}.
Recently there has been much activity connected with tests of this model,
particularly on measuring the unitarity triangle through the time evolution
measurements\cite{GL,LNQS,FWX} and time-independent measurements of
decay rates by using SU(3) relations for B-meson Decay
-amplitudes \cite{SW,GRL}.
Other models for CP violation, like the superweak theory\cite{LW},
or the most general two higgs doublet models\cite{WW},
may have different predictions for the
direct and indirect phases. In the K-system, several experiments are
under way to probe direct CP violation and time evolution of kaon decays.
In the D-system, mixing and CP violation are expected to be small in the
standard model (SM) and thus experimental tests are interesting as a probe
of new physics. In the B-system, CP violation could be large in the CKM
scheme of the SM, prompting much experimental activity.

  Recently, it has been pointed out \cite{DH,BF} that
there are some limitations in extracting the angles of the unitarity triangle
by using SU(3) relations for B-meson decay amplitudes as suggested by
\cite{GRL}. Thus it is still necessary to
investigate carefully the time-evolution measurements.
Basic formula for time-dependent decay rates have been extensively
studied\cite{BS,WOLF,PZ,DR,BKUS,GL,LNQS,FWX} and
applied to various processes. In this note we develop and refine these
studies into a   general model-independent and
rephase-invariant formalism that cleanly relates observables to the fundamental
parameters and explicitly separates different forms of CP violation for the
neutral meson systems.
A meaningful classification of different forms of CP violation must be
invariant against phases \cite{JARLSKOG}
that can be arbitrarily assigned. In this note we will show that
in the neutral meson system  there exist in general seven
 rephase-invariant observables which in principle
can be detected by studying the
time-evolution of the neutral meson and rate asymmetry.
We then conclude that  CP-violating observables are in general
classified into three types of CP violation  and any CP-violating observable
can be expressed in terms of
 seven rephase-invariant quantities.
Especially, we emphasize  the
importance of  the rephase-invariant and
model-independent  observables when interpreting experimental measurements
in the $K^0$-, $D^0$- and $B^0$-systems.  Applying this analysis
to the decays $B^0 \rightarrow \pi^+ \pi^-$, $\pi^0 \pi^0$ and
$B^+ \rightarrow \pi^+ \pi^0$, we find there are eight physical observables
to determine nine input parameters when including the electroweak penguin.
Therefore, to  extract the angle $\alpha$ of the unitarity triangle,
one needs in principle an additional  theoretical input.  We also examine
an interesting  decay mode to extract the angle $\gamma$.

   Let $M^0$ be the neutral meson (which can be $K^0$ or $D^0$ or $B^0$)
and $\bar{M}^0$ its antiparticle. $M^0$
and $\bar{M^0}$ can mix with each other and form two physical mass eigenstates

\begin{equation}
M_1  =  p| M^0 > + q | \bar{M^0} >, \qquad
 M_2  =  p| M^0 > - q | \bar{M^0} >
\end{equation}
The CP-violating parameter $\epsilon_{M}$ is introduced via
\begin{equation}
\epsilon_{M} = \frac{1-q/p}{1 + q/p} \  , \qquad \frac{q}{p} \equiv \sqrt{
\frac{H_{21}}{H_{12}}}
\end{equation}
where $H_{12} \equiv M_{12} - \frac{i}{2} \Gamma_{12} =
< M^0 | H_{eff} |\bar{M}^0 > $.

  Let $f$ denote the final decay state of the neutral meson and $\bar{f}$
its charge conjugate state. The decay amplitudes of $M^0$ and $\bar{M^0}$
are denoted by
\begin{equation}
g \equiv <f|H_{eff}| M^0>, \   h  \equiv  <f|H_{eff}|\bar{M^0} >; \
\bar{g}  \equiv  <\bar{f}|H_{eff}|\bar{M^0}>, \
\bar{h}  \equiv  <\bar{f}|H_{eff}|M^0 >
\end{equation}
Parameters containing direct CP violation are defined by

\begin{equation}
\epsilon_{M}'  \equiv  \frac{1-h/g}{1+h/g}, \
 \bar{\epsilon}_{M}' \equiv \frac{1-\bar{g}/\bar{h}}{1+ \bar{g}/\bar{h}}; \
 \epsilon_{M}''  \equiv  \frac{1-\bar{g}/g}{1+ \bar{g}/g}, \   \bar{
\epsilon}_{M}'' \equiv \frac{1- h/\bar{h}}{1+ h/\bar{h}}
\end{equation}
Note that the above
parameters are not physical observables
since they are not rephase-invariant.  Let us introduce
 CP-violating observables by  considering the ratio, \mbox{$\eta_{f} \equiv
<f |H_{eff} | M_{2} > /< f |H_{eff} | M_{1} >
= (1 - r_{f})/(1 + r_{f})$},
with $r_{f} = (q/p)(h/g)$ being rephase-invariant.  Using a simple
algebra relation $1-ab = [(1+a)(1-b) + (1-a)(1+b)]/2$ and $1+ab = [(1+a)(1+
b) + (1-a)(1-b)]/2$, it is not difficult to show that $\eta_{f}$ can be
rewritten as
\begin{equation}
\eta_{f} = \frac{a_{\epsilon} + a_{\epsilon'} + i\ a_{\epsilon +
\epsilon'}}{2 + a_{\epsilon} a_{\epsilon'} + a_{\epsilon \epsilon'}}
\end{equation}
with

\begin{eqnarray}
a_{\epsilon} & = &  \frac{1 - |q/p|^2}{1 + |q/p|^2} =
\frac{2 Re \epsilon_{M}}{1 + |\epsilon_{M}|^2} \ ,  \qquad
a_{\epsilon'} =  \frac{1 - |h/g|^2}{1 + |h/g|^2} =
\frac{2 Re \epsilon_{M}'}{1 + |\epsilon_{M}'|^2} \ ; \nonumber \\
a_{\epsilon + \epsilon'} & = & \frac{-4 Im(qh/pg)}{(1+|q/p|^2)(1+|h/g|^2)}
= \frac{2 Im\epsilon_{M} (1-|\epsilon_{M}'|^2) + 2 Im\epsilon_{M}'
(1-|\epsilon_{M}|^2) }{ (1 + |\epsilon_{M}|^2 )(1+|\epsilon_{M}'|^2)}  \\
a_{\epsilon \epsilon'} & = & \frac{4 Re(qh/pg)}{(1+|q/p|^2)(1+|h/g|^2)} -1
= \frac{4 Im\epsilon_{M} \  Im \epsilon_{M}' - 2
(|\epsilon_{M}|^2 + |\epsilon_{M}'|^2) }{( 1 + |\epsilon_{M}|^2)
( 1 + |\epsilon_{M}'|^2)} \nonumber
\end{eqnarray}
Obviously, $a_{\epsilon}$, $a_{\epsilon'}$, $a_{\epsilon + \epsilon'}$ and
$a_{\epsilon \epsilon'}$ are all rephase-invariant.
But only three of them are independent as
 $(1- a_{\epsilon}^{2})(1 - a_{\epsilon'}^{2}) =  a_{\epsilon + \epsilon'}^{2}
 + (1+ a_{\epsilon \epsilon'})^{2}$. Analogously, one has
\begin{equation}
\eta_{\bar{f}} \equiv \frac{<\bar{f} |H_{eff} | M_{2} > }{<\bar{f} |H_{eff} |
M_{1} > } = \frac{a_{\epsilon} + a_{\bar{\epsilon}'} + i\ a_{\epsilon +
\bar{\epsilon}'}}{2 + a_{\epsilon} a_{\bar{\epsilon}'} +
a_{\epsilon \bar{\epsilon}'}}
\end{equation}
where $a_{\bar{\epsilon}'}$, $a_{\epsilon + \bar{\epsilon}'}$ and
$a_{\epsilon \bar{\epsilon}'}$ are similar to
$a_{\epsilon'}$, $a_{\epsilon + \epsilon'}$ and
$a_{\epsilon \epsilon'}$ but with $\epsilon_{M}'$ being replaced by
$\bar{\epsilon}_{M}'$.

   In most phase conventions of the CKM matrix  in the
literature\cite{CKM,WOLF2}, one has, $|\epsilon_{K}| \ll 1$, for the K-system.
In the phase convention
of Wu and Yang\cite{WY}, one has, further, $|\epsilon'_{K}| \ll 1$.
With the fact that $\omega = |A_{2}/A_{0}| \ll 1$,
one obtains from eq.(6) that $ a_{\epsilon} \simeq 2 Re \epsilon_{K}$,
$a_{\epsilon'} \simeq 2 Re\epsilon'_{K}$, $a_{\epsilon + \epsilon'}
\simeq 2 Im \epsilon_{K} + 2 Im \epsilon'_{K}$ and $a_{\epsilon \epsilon'}
\simeq 0$, and thus

\begin{equation}
\eta_{+-} \simeq Re \epsilon_{K} + Re \epsilon_{K}' + i (
Im \epsilon_{K} + Im \epsilon_{K}' ) = \epsilon_{K} + \epsilon'_{K}
\end{equation}
which reproduces the form often used in the
literature for $K^0 \rightarrow \pi^+ \pi^- $ decay.

  Two additional rephase-invariant quantities complete the set of
observables,

\begin{equation}
a_{\epsilon''} = \frac{1-|\bar{g}/g|^{2}}{1 + |\bar{g}/g|^2 }
= \frac{2 Re \epsilon_{M}'' }{1 + |\epsilon_{M}''|^2 }, \qquad
a_{\bar{\epsilon}_{M}''} =  \frac{1-|\bar{h}/h|^{2}}{1 +
|\bar{h}/h|^2 } = \frac{2 Re \bar{\epsilon}_{M}'' }{1 +
|\bar{\epsilon}_{M}''|^2 }
\end{equation}

    So far we have introduced five parameters from which
 seven independent  rephase-invariant
observables are constructed to describe CP violation: $\epsilon_{M}$
is an indirect CP-violating parameter;
$\epsilon_{M}''$  and $\bar{\epsilon}_{M}''$
define direct CP-violating parameters; $\epsilon_{M}'$ and
$\bar{\epsilon}_{M}'$  contain the ratio of the two decay amplitudes and
can be associated with direct CP violation as well as the interference
between indirect and direct CP violation.  All the CP
violations can be well defined and in general classified into
the following three types: i) purely indirect CP violation which  is given by
the  rephase-invariant CP-violating observable  $a_{\epsilon}$;
ii) purely direct CP violation which is characterized by the
rephase-invariant CP-violating observables $a_{\epsilon''}$ and $a_{\bar{
\epsilon}''}$; and iii) indirect-direct mixed CP violation which
is described by the rephase-invariant CP-violating observables
$a_{\epsilon + \epsilon'}$ and $a_{\epsilon + \bar{\epsilon}'}$.
For the case that the final states are CP eigenstates, one has
$a_{\epsilon'} = a_{\epsilon''}= a_{\bar{\epsilon}'}= a_{\bar{\epsilon}''}$
. Thus, in this case $a_{\epsilon'}$ and $ a_{\bar{\epsilon}'} $
also indicate  purely direct CP violation.  When the final states are
not CP eigenstates, $a_{\epsilon'}$ and $ a_{\bar{\epsilon}'} $
do not,  in general, provide a clear signal of direct CP violation
although they contain direct CP violation. Their
deviation from $a_{\epsilon'} = \pm 1, \  0$ and $ a_{\bar{\epsilon}'}= \mp 1,
 \  0 $ can arise from different CKM angles,  final state interactions, or
different hadronic form factors, but not necessarily from CP violation.

    In order to measure these rephase-invariant observables,
we consider the proper time evolution of the neutral mesons

 \begin{equation}
|M^{0}(t) >  =  \sum_{i=1}^{2} C_{i} e^{-i(m_{i} - i \Gamma_{i}/2)t }
|M_{i} >  \ ; \qquad
|\bar{M^{0}}(t) > =  \sum_{i=1}^{2} \bar{C_{i}} e^{-i(m_{i} -
i \Gamma_{i}/2)t } |M_{i} >
\end{equation}
with $C_{1} = C_{2} = 1/2p $ and $\bar{C_{1}} = -
\bar{C_{2}} = 1/2q$ for purely $M^0$ and $\bar{M}^0$ at $t=0$.
The time-dependent decay rates are found to be

\begin{eqnarray}
 & \Gamma &( M^{0} (t) \rightarrow f ) \propto |<f |H_{eff} | M^{0} (t) >|^2
= \frac{1}{1 + a_{\epsilon}} \frac{(|g|^2 + |h|^2)}{2} e^{-\Gamma t} \\
& & \cdot [ (1+ a_{\epsilon} a_{\epsilon'} ) \cosh (\Delta\Gamma t)
+ (1+a_{\epsilon \epsilon'}) \sinh (\Delta\Gamma t) + (a_{\epsilon} + a_{
\epsilon'} ) \cos (\Delta m t) + a_{\epsilon + \epsilon'} \sin (\Delta m t) ]
\nonumber \\
& \Gamma & ( \bar{M^{0}} (t) \rightarrow \bar{f} )
 \propto  |<\bar{f} |H_{eff} | \bar{M^0} (t) >|^2
= \frac{1}{1- a_{\epsilon}} \frac{(|\bar{g}|^2 + |\bar{h}|^2)}{2}
 e^{-\Gamma t} \\
& & \cdot [ (1+ a_{\epsilon} a_{\bar{\epsilon}'} ) \cosh (\Delta\Gamma t)
+(1+ a_{\epsilon \bar{\epsilon}'}) \sinh (\Delta\Gamma t) - (a_{\epsilon} + a_{
\bar{\epsilon}'} ) \cos (\Delta m t)  -  a_{\epsilon + \bar{\epsilon}'}
\sin (\Delta m t) \  ] \nonumber
\end{eqnarray}
One can easily  write down the decay rates $\Gamma (M^{0}(t) \rightarrow
\bar{f} )$ and $\Gamma (\overline{M}^{0}(t) \rightarrow f )$.
  where $\Delta \Gamma = \Gamma_2 - \Gamma_1 $ and $\Delta m = m_2 - m_1 $.
Here we have omitted the integral of the phase space.

 From studying the time-dependent spectrum of the decay rates of $M^0$ and
$\bar{M^0}$, one can, in principle, find the coefficients of the
four functions $\sinh (\Delta \Gamma t) $, $\cosh (\Delta \Gamma t) $ ,
$\cos (\Delta m t) $ and $\sin (\Delta m t) $ and extract the quantities
$a_{\epsilon}$,
$a_{\epsilon'}$, $a_{\epsilon + \epsilon'}$, $|g|^2 + |h|^2 $,
$a_{\bar{\epsilon}'}$, $a_{\epsilon + \bar{\epsilon}'}$ and $|\bar{g}|^2 +
|\bar{h}|^2 $ as well as $\Delta m $, $\Delta \Gamma $ and $\Gamma$. Thus,
one can determine the amplitudes $|q/p|$, $|g|^2$, $|h|^2$, $|\bar{g}|^2$,
$|\bar{h}|^2$ and combinations of the phases $(\phi_{M} + \phi_{A})$ as
well as $(\phi_{M} + \bar{\phi}_{A})$ \footnote{ Note that only the combination
of the two phases is rephase-invariant.} via

\begin{equation}
|\frac{q}{p}|^{2} = \frac{1- a_{\epsilon}}{1 + a_{\epsilon}}, \qquad
|\frac{h}{g}|^{2} = \frac{1 - a_{\epsilon'}}{1 + a_{\epsilon'}}, \qquad
\sin (2(\phi_{M} + \phi_{A} )) = \frac{a_{\epsilon + \epsilon'}}{
\sqrt{(1-a_{\epsilon}^{2})(1-a_{\epsilon'}^{2})} }
\end{equation}
where $q/p = |q/p|e^{-2i\phi_{M}}$, $h/g = |h/g| e^{-2i\phi_{A}}$ and
$\bar{g}/\bar{h} = |\bar{g}/\bar{h}| e^{-2i\bar{\phi}_{A}}$.
For $f^{CP} = \pm f$, one has $\phi_{A} = \bar{\phi}_{A}$. For $f \not =
f^{CP}$, $\phi_{A} = \bar{\phi}_{A}$ holds only when final state
interactions are absent,
so that $\phi_{A} \neq \bar{\phi}_{A}$ implies an existence of  final
state interactions when $f$ is not a CP eigenstate.

  The time-dependent CP asymmetry is  the difference between
the two decay rates of eqs. (11) and (12). In terms of the
rephase-invariant quantities, we have
\begin{eqnarray}
\Delta_{CP}(t) & = & \Gamma(M^0 (t) \rightarrow f) - \Gamma (\overline{M}^0
(t) \rightarrow \bar{f} ) = \frac{1}{1 - a_{\epsilon}^{2} } \frac{1}{2}
e^{-\Gamma t}\frac{(|\bar{g}|^{2} + |g|^2 )}{2} \cdot  \nonumber \\
& &  \{ [ (1+a_{\epsilon''}) (1 -  a_{\epsilon}) (a_{\epsilon}
+ a_{\epsilon'} ) + (1-a_{\epsilon''}) (1 +  a_{\epsilon})
(a_{\epsilon} + a_{\bar{\epsilon}'} ) ] \cos (\Delta m t) \nonumber \\
 & & + [ (1+a_{\epsilon''})  (1 -  a_{\epsilon}) a_{\epsilon +
\epsilon'}  + (1-a_{\epsilon''}) (1 +  a_{\epsilon})
a_{\epsilon + \bar{\epsilon}'}  ] \sin (\Delta m t) \   \\
 & & -[(a_{\epsilon} - a_{\epsilon''})
(2 + (a_{\bar{\epsilon}'} + a_{\epsilon'} ) a_{\epsilon} ) + (1-
a_{\epsilon} a_{\epsilon''} )(a_{\bar{\epsilon}'} - a_{\epsilon'})
a_{\epsilon}  ]  \cosh (\Delta \Gamma t)    \nonumber \\
& & - [(a_{\epsilon} - a_{\epsilon''})
(2 + a_{\epsilon \bar{\epsilon}'} + a_{\epsilon \epsilon'} ) +
  (1-  a_{\epsilon} a_{\epsilon''} )(a_{\epsilon \bar{\epsilon}'} -
a_{\epsilon \epsilon'}  )] \sinh (\Delta \Gamma t) \}  \nonumber \\
& & + ( g \leftrightarrow h\ , \bar{g} \leftrightarrow \bar{h}\ ,
\  a_{\epsilon''}
\leftrightarrow - a_{\bar{\epsilon}''} )  \nonumber
\end{eqnarray}

 One can in general define several asymmetries from the four time-dependent
decay rates $\Gamma (M^{0}(t) \rightarrow f )$,
$\Gamma (\overline{M}^{0}(t) \rightarrow \bar{f}  )$ ,
$\Gamma (M^{0}(t) \rightarrow  \bar{f} )$ and
$\Gamma (\overline{M}^{0}(t) \rightarrow f )$ .

 To apply the above general analyses to specific processes, we may classify
the processes
into the following scenarios

 i) \  $M^{0} \rightarrow f $ ($M^{0} \not\rightarrow \bar{f} $) , \
$\overline{M}^0 \rightarrow \bar{f}$ ($\overline{M}^0 \not \rightarrow f$)
, i.e., $f$ or $\bar{f}$ is not a common final
state of $M^{0}$ and $\overline{M}^{0}$.
Examples are:\  $M^0 \rightarrow M'^- \bar{l} \nu $, $\bar{M}^{0}
\rightarrow M'^+ l \bar{
\nu} $; \  $B^0 \rightarrow D^- D_{s}^{+} $; \
$\bar{B}^{0} \rightarrow D^+ D_{s}^{-} $.
,
This scenario also applies to charged meson decays\cite{KP}.

 ii) \  $M^{0} \rightarrow (f = \bar{f}, \  f^{CP} =  f) \leftarrow
 \overline{M}^0$, i.e., final states are CP eigenstates.
Such as $B^{0} (\bar{B}^{0} )$, $D^{0} (\bar{D}^{0} )$,
$K^{0} (\bar{K}^{0} )$  $\rightarrow \pi^{+} \pi^{-} $, $\pi^{0} \pi^{0}$,
\ $\cdots $. For the final states such as $\pi^- \rho^+$ and $\pi^+ \rho^-$
, although each of them is not a CP eigenstate of $B^{0}(\bar{B}^{0})$ or
$D^0 (\bar{D}^0)$, one can always reconstruct them into CP eigenstates
as $(\pi \rho)_{\pm} = (\pi^- \rho^+ \pm \pi^+ \rho^- )$ with
$CP (\pi \rho)_{\pm} = \pm (\pi \rho)_{\pm}$. This reconstruction is
meaningful since $\pi^- \rho^+$ and  $\pi^+ \rho^-$ have the same weak
phase as they contain the same quark content.

 iii) \  $M^{0} \rightarrow ( f, \  f \not \rightarrow f^{CP})  \leftarrow
 \overline{M}^0$, i.e., final states are common final states but they are
not charge conjugate states. For example,
$B^{0} (\bar{B}^{0} )
\rightarrow K_{S} J/\psi $,
$B^{0}_{s} (\bar{B}^{0}_{s} ) \rightarrow K_{S} \phi $.

 iv)  \  $M^{0} \rightarrow (f \  \& \  \bar{f}, \  f^{CP} \neq f ) \leftarrow
 \overline{M}^0$ , i.e., both $f$ and $\bar{f}$ are the common final
states of $M^0$ and $\overline{M}^0$, but they are not CP eigenstates.
This is the most general case. For example, \
$B^{0} (\bar{B}^{0} ) \rightarrow D^- \pi^{+}$, $\pi^{-} D^{+}$ ; \
$D^{-} \rho^{+}$, $\rho^{-} D^{+}$;\
$B^{0}_{s} (\bar{B}^{0}_{s} ) \rightarrow D_{s}^{-} K^{+}$,  $K^{-} D_{s}^{
+}$.

   In the scenario i),    one has: $a_{\epsilon'} = - a_{\bar{\epsilon}'} = 1$
, $ a_{\epsilon + \epsilon'} = 0  = a_{\epsilon + \bar{\epsilon}'} $ and
$a_{\epsilon \epsilon'} = -1 = a_{\epsilon \bar{\epsilon}'}$. The
time-dependent rates of eqs. (11) and (12) then become much simpler.
Thus,  $\Delta m $, $\Delta \Gamma $, $a_{\epsilon}$ and $a_{\epsilon''}$
can be easily extracted via

\begin{eqnarray}
& & A_{CP}(t)  =  \Delta_{CP}(t)/(\Gamma(M^{0}(t) \rightarrow f) + \Gamma
(\overline{M}^{0} (t) \rightarrow \bar{f} ))  = a_{\epsilon''}  \\
& & A'_{CP} (t) =  \frac{\Gamma(\overline{M}^{0}(t) \rightarrow f) -
\Gamma (M^{0}(t) \rightarrow \bar{f} )}{\Gamma(\overline{M}^{0}(t)
\rightarrow f) + \Gamma (M^{0} (t) \rightarrow \bar{f} )}
= (a_{\epsilon''} + \frac{2 a_{\epsilon}}{1 + a_{\epsilon}^{2}})/(1 +
\frac{2 a_{\epsilon}}{1 + a_{\epsilon}^{2}}a_{\epsilon''}) \\
& & \frac{\cos (\Delta m t)}{\cosh (\Delta \Gamma t )}  =
\frac{A^{f}(t) + a_{\epsilon}}{1 + A^{f}(t) a_{\epsilon} }; \qquad
A^{f}(t) = \frac{\Gamma(M^{0}(t) \rightarrow f) - \Gamma (\overline{M}^{0}(t)
\rightarrow f )}{\Gamma(M^{0}(t) \rightarrow f) + \Gamma
(\overline{M}^{0} (t) \rightarrow f )} .
\end{eqnarray}
It is interesting to note that the CP asymmetries $A_{CP}$ and $A'_{CP}$
are actually time-independent since the time-
dependent parts cancel in the ratio. Indirect CP violation
$a_{\epsilon}$ in the $B^{0}$- and $B^{0}_{s}$-systems may be
directly obtained by measuring the decay channels in the scenario i),
such as $B^0 \rightarrow D^- K^+$, $\pi^- D_{s}^{+}$ and
$B_{s}^{0} \rightarrow D_{s}^{-} \pi^+$, $K^{-} D^+$, respectively, or
from their semileptonic decays,
 since in these decays one has $a_{\epsilon''}=0$.  Purely direct CP violation
$a_{\epsilon''}$ in the $B^{0}$- and $B^{0}_{s}$-systems may be detected
by studying decay modes such as  $B^0 \rightarrow D^{-} D^{+}_{s}$
, $\pi^{-} K^{+}$, and  $B_{s}^{0} \rightarrow D_{s}^{-} D^{+}$, $K^{-}
\pi^+$.   These decay modes receive contributions from both tree and
penguin diagrams so that the final state interactions may become significant.

  We now  discuss the scenario ii) in which
$a_{\epsilon'} = a_{\epsilon''}= a_{\bar{\epsilon'}}= a_{\bar{\epsilon}''}$
and  $a_{\epsilon + \epsilon'} = a_{\epsilon + \bar{\epsilon}'}$.  Thus,
the time-dependent CP asymmetry simplifies to

\begin{eqnarray}
& & A_{CP}(t)  =  (\Delta_{m}(t) - a_{\epsilon}\
\Delta_{\gamma}(t))/(\Delta_{\gamma}(t) - a_{\epsilon}\  \Delta_{m}(t))
\nonumber \\
& & \Delta_{m}(t) = (a_{\epsilon} + a_{\epsilon'} ) \cos (\Delta m t) +
a_{\epsilon + \epsilon'}   \sin (\Delta m t)  \\
& & \Delta_{\gamma}(t) = (1 + a_{\epsilon} a_{\epsilon'} )
\cosh (\Delta \Gamma t)  + (1 + a_{\epsilon \epsilon'} )
 \sinh (\Delta \Gamma t)  \nonumber
\end{eqnarray}

 Suppose that $\Delta m$, $\Delta \Gamma $ and $a_{\epsilon}$ are known
from studying the processes in the scenario i); one then can extract the
direct CP-violating observable $a_{\epsilon'}$ and CP-violating observable
$a_{\epsilon + \epsilon'}$ from the coefficients of $\cos (\Delta m t)$ and
$\sin (\Delta m t)$ respectively from type (ii) processes.

  Let us consider the following special but realistic cases:

  1)\  $a_{\epsilon} \ll 1$,  then, to the first
order of $a_{\epsilon}$ and $a_{\epsilon'}$, one has

\begin{equation}
A_{CP}(t) \simeq  - a_{\epsilon} +
 \frac{  (a_{\epsilon} + a_{\epsilon'} ) \cos (\Delta m t) +
a_{\epsilon + \epsilon'}   \sin (\Delta m t) }{ \cosh (\Delta \Gamma t)  +
(1 + a_{\epsilon \epsilon'})  \sinh (\Delta \Gamma t)}
\end{equation}
This case actually holds for all the neutral meson systems.
In the $K^{0}-\bar{K}^{0}$ system, one also has $a_{\epsilon'} \ll 1$
and $\Delta m \simeq - \Delta \Gamma /2$.

 2)\   $a_{\epsilon} \ll 1$,  $|\Delta \Gamma |
\ll |\Delta m |$ and $|\Delta \Gamma /\Gamma | \ll 1$. Then, $A_{CP}(t)$
further simplifies

\begin{equation}
A_{CP}(t) \simeq  - a_{\epsilon} +
 (a_{\epsilon} + a_{\epsilon'} ) \cos (\Delta m t) +
a_{\epsilon + \epsilon'}   \sin (\Delta m t)
\end{equation}
which may be applied,  in a good approximation, to the $B^{0} - \bar{B}^{0}$
system.

  As an interesting example, let us reanalyze the $\pi \pi$ processes in
the context of our formalism and
show how eight observables may be extracted from the data.

   The decay amplitudes of $B^{0} \rightarrow \pi^+ \pi^-$, $\pi^0 \pi^0$
and  $B^{+} \rightarrow \pi^+ \pi^0$ can be expanded, in terms of the
isospin ($I=0,2$) and the tree- and penguin-diagrams, into

\begin{eqnarray}
g_{+-} & = & \sqrt{2} (A_{0T}e^{-i\phi_{T} +i\delta_{0T}} + A_{0P}e^{-i\phi_{P}
+i\delta_{0P}} - A_{2T}e^{-i\phi_{T} +i \delta_{2T}}-
A_{2P}e^{-i\phi_{P} +i \delta_{2P}}) \\
g_{00} & = & A_{0T}e^{-i\phi_{T} +i\delta_{0T}} + A_{0P}e^{-i\phi_{P}
+i\delta_{0P}} +2 (A_{2T}e^{-i\phi_{T} +i \delta_{2T}}
+A_{2P}e^{-i\phi_{P} +i \delta_{2P}}) \\
g_{+0} & = & 3 (A_{2T}e^{-i\phi_{T} +i \delta_{2T}} +
A_{2P}e^{-i\phi_{P} +i \delta_{2P}})
\end{eqnarray}
where $A_{0i}$ and $A_{2i}$ are the isospin $I=0$ and 2 amplitudes of the
tree diagram ($i=T$) and penguin diagrams ($i=P$), $\delta_{0i}$ and
$\delta_{2i}$ ($i=T$, $P$) are the corresponding strong
phases. Where $A_{2P}$ arises from the electroweak penguin.
$\phi_{T}$ and $\phi_{P}$ are the weak phases of the tree- and penguin-
diagrams respectively. Thus, we have nine physical
quantities. They consist of five phases
and four amplitudes:
\begin{eqnarray}
& & \Delta \phi = \phi_{T} - \phi_{P}, \   (\phi_{M} + \phi_{T}) \ or \
(\phi_{M} + \phi_{P}), \  \Delta \delta_{T} = \delta_{0T} - \delta_{2T},
\nonumber \\
  & & \Delta \delta_{P} = \delta_{0P} - \delta_{2T}, \
\Delta \delta_{P}' = \delta_{2P} - \delta_{2T},  \  A_{0T}, \
A_{2T}, \  A_{0P}, \  A_{2P}
\end{eqnarray}
while we have in general eight independent observables, they are:

\begin{equation}
a_{\epsilon'}^{(+-)}, \ a_{\epsilon'}^{(00)}, \
\ a_{\epsilon''}^{(+0)}, \ a_{\epsilon + \epsilon'}^{(+-
)}, \  a_{\epsilon + \epsilon'}^{(00)},  \ <\Gamma_{+-}>, \
<\Gamma_{00}>, \  <\Gamma_{+0}>
\end{equation}
which implies that without using a theoretical input one cannot
in principle unambiquously  extract
the angle $\alpha$ from $B\rightarrow \pi\pi$.
A model calculation however
shows that $A_{2p}/A_{2T} \simeq 1.6\% (|V_{td}|/|V_{ub}|
)$ \cite{DH}. If the contribution of the electroweak penguin is negligible,
it is remarkable that
there are only seven quantities (omitting $\Delta \delta_{P}'$ and $A_{2P}$
in eq. (24)) which can be extracted from seven observables (omitting
$a_{\epsilon''}^{(+0)}$ in eq. (25)),
where $a_{\epsilon'}^{(+-)}$ and $a_{\epsilon'}^{(00)}$ are proportional to
$\sin (\Delta\delta_{P}) \sin(\Delta \phi) $ and $\sin (\Delta\delta_{T}
-\Delta \delta_{P}) \sin(\Delta \phi) $;\
$a_{\epsilon + \epsilon'}^{(+-)}$ and $a_{\epsilon + \epsilon'}^{(00)}$
are related to combinations of $\sin(\phi_{M} + \phi_{T})$,
$\cos (\Delta \delta_{P})$, $\cos (\Delta \delta_{T})$ and
the relative amplitude of the tree and penguin diagrams.
Together with the decay rates
$\Gamma_{+-}$, $\Gamma_{00}$ and $\Gamma_{+0}$, one can in principle extract
 the  seven quantities in eq. (24) as also noted by \cite{GL}.
In the CKM scheme, one has $\phi_{M} = \beta$,  $\phi_{T} =
\gamma$ and $\phi_{P} \simeq -\beta$, thus
\begin{equation}
(\phi_{M} + \phi_{T}) = \beta + \gamma = \pi - \alpha, \qquad
\Delta \phi = \phi_{T} - \phi_{P} \simeq \beta + \gamma = \pi - \alpha
\end{equation}

 A model for the strong phase of the strong penguin diagram
has recently been studied by \cite{KP,SDW}.
As an illustration of the effect of strong phase and amplitude of the
strong penguin diagram,   we present in the table 1
the values of $a_{\epsilon'}^{(+-)}$,  $a_{\epsilon'}^{(00)}$,
$a_{\epsilon + \epsilon'}^{(+-)}$, and $a_{\epsilon + \epsilon'}^{(00)}$
with $a_{\epsilon}=0$ in the decays $B^{0} \rightarrow \pi^+ \pi^-$,
$\pi^0 \pi^0$ as functions of representative Wolfenstein parameters
$\rho$ and $\eta$
of the CKM matrix in a BSW model\cite{BSW} using the methods and parameters
of refs. \cite{KP,AL}.   As we see from table 1, the influence of the
amplitude and strong phase of the penguin diagram is sizable even in the
$B^0 \rightarrow \pi^+ \pi^-$ channel but is quite drastic in the
$B^0 \rightarrow \pi^0 \pi^0$ channel, where color suppression occurs for
the tree amplitude.  It should also be noted how sensitive
$a_{\epsilon + \epsilon'}^{(00)}$ is
to the CKM parameters, changing sign between the
two prefered Ali and London\cite{AL}  solutions
due to the cancellation between tree
and strong penguin contributions. For certain values of $\rho$ and $\eta$,
$a_{\epsilon + \epsilon'}^{(00)}\simeq 0$. In this case the electroweak
penguin could be a complication in the $\pi^0 \pi^0$ channel
when extracting the angle $\alpha$.
The effect of the electroweak penguin diagram may be seen
by studying the CP asymmetry $a_{\epsilon''}^{(+0)}$ in the charged B-meson
decay.

  A similar analysis can be applied to the K- and D-systems. It is of
interest that for the K-system it becomes simpler, i.e.
 $\delta_{IT} = \delta_{IP} = \delta_{I}$
($I=0$,$2$). This is because in the K-system
$\pi \pi$ scattering is assumed to be purely elastic (in the good
approximation of neglecting the $\pi \pi \gamma$ final state),
while in the B-system $\pi \pi$ scattering
can be inelastic. Thus in the K-system
there are only seven parameters in eq. (24) instead of nine,
 but there are also only six independent observables in eq. (25)
because  $\Delta = [\Gamma (\pi^+ \pi^-) -
\bar{\Gamma}(\pi^+ \pi^-) ] + [\Gamma (\pi^0 \pi^0) -
\bar{\Gamma}(\pi^0 \pi^0) ] = 0 $ as required by CPT in the absence of
other channels. In the B-system there are two more parameters as well as
two more observables as shown above and $\Delta \neq 0$. Therefore
the measurement of $\Delta$ distinguishes CP violation
in  the K-system from that of B-system \footnote{ We
would like to thank L. Wolfenstein for pointing this out to us}.

 We now present another
 interesting decay mode to extract the angle $\gamma$, i.e.,
$B_{s}^{0} \rightarrow ( D_{s}^{-}K^{+}, \  K^{-}D_{s}^{+}) \leftarrow
\bar{B}_{s}^{0} $.
 Since these decay modes only receive contributions from tree diagram, the
phase $\phi_{A}$ of the amplitude is almost a purely weak phase and given
by $2\phi_{A} = \gamma$ in the CKM scheme.  The phase $\phi_{M}$ from
the $B_{s}^{0}-\bar{B}_{s}^{0}$ mixing is expected to be small in the CKM
scheme, $\phi_{M} \ll 1$. By measuring the rephase-invariant quantities
$a_{\epsilon + \epsilon'}$ and $a_{\epsilon'}$  in the above decay modes,
one then extracts the angle $\gamma$ (see equation (13))
\begin{equation}
\sin (2(\phi_{M} + \phi_{A})) \simeq \sin\gamma  = \frac{a_{\epsilon +
\epsilon'}}{\sqrt{(1-a_{\epsilon}^{2})(1- a_{\epsilon'}^{2})}}
\simeq \frac{a_{\epsilon +
\epsilon'}}{\sqrt{(1- a_{\epsilon'}^{2})}}
\end{equation}
In general,  such a triangle is not necessarily  closed as long
as either the mass mixing matrices or the amplitudes receive
contributions from new sources of CP violation or new interactions beyond
the standard model,  such as the most general two-Higgs
doublet model\cite{WW}.

  We hope that the general model-independent and rephase-invariant
formalism developed in this note will be useful in the analysis of the
neutral meson systems to test the SM and probe new physics.

 The authors would like to thank G. Kramer and L. Wolfenstein for carefully
reading the manuscript and for helpful advice. They also wish to thank H. Simma
for useful discussions.

\newpage

\def\head{\vspace{-7cm}\begin{flushright} DESY 93-192\\[-2mm]
                                          December 1993\\[1cm]
                        \end{flushright}}
\setlength{\topmargin}{1pt}
\setlength{\textheight}{604pt}
\setlength{\oddsidemargin}{1pt}
\setlength{\evensidemargin}{1pt}
\setlength{\textwidth}{446pt}
\renewcommand{\textfraction}{.0}
\renewcommand{\topfraction}{1.0}
\renewcommand{\bottomfraction}{1.0}

 Table 1. $B\rightarrow \pi \pi $ CP-violating observables in a standard
model calculation using the BSW model \cite{BSW}, absorptive parts of the
penguins in NLL formulism \cite{SDW}, and the CKM parameters of a recent
fit \cite{AL}. \\

\newcommand{\ba}{\begin{array}}
\newcommand{\ea}{\end{array}}
\newcommand{\bd}{\begin{displaymath}}
\newcommand{\ed}{\end{displaymath}}
\newcommand{\be}{\begin{equation}}
\newcommand{\ee}{\end{equation}}
\newcommand{\bea}{\begin{eqnarray}}
\newcommand{\eea}{\end{eqnarray}}

\def\nc{\\}
\def\sc{\\[-1mm]}

\def\empty{ (---) &&&&& }
\def\same{ }

\small
\begin{tabular}{||l||l|l||l|l|l||}
\hline\hline
\multicolumn{6}{||c||}{{CP Violation in $B \to \pi\pi$}}\\
\multicolumn{6}{||c||}{Tree (T) and Penguins with (P) or without (P')
				Absorptive Parts} \\
\multicolumn{6}{||c||}{{NLL QCD Coefficients, BSW model}}\\
\hline
           &\multicolumn{2}{|c||}{\bf{Assumptions}}
                            &\multicolumn{3}{|c||}{\bf{Decay Parameters}}\\
\hline
Channel
&CKM ($\rho,\eta$)&Amplitude&$a_{\epsilon'}$&$a_{\epsilon+\epsilon'}$&$<BR>$\\
\hline \hline
           & (-0.12,0.34)     &T + P    &0.316   &0.201
 &$4.25\times 10^{-7}$\\
$\pi^0\pi^0$& ~~~~~~~" &T + P'   &0.0     &0.216     &$4.14\times 10^{-7}$\\
           & ~~~~~~~"     &T        & 0.0    &-0.951    &$5.66\times 10^{-7}$
\\
\hline
           & (-0.28,0.24)     &T + P    &0.496   &-0.465
 &$1.91\times 10^{-7}$\\
$\pi^0\pi^0$& ~~~~~~~" &T + P'   &0.0     &-0.487    &$1.77\times 10^{-7}$\\
           & ~~~~~~~"     &T        &0.      &-0.869    &$5.93\times 10^{-7}$\\
\hline
           & (-0.12,0.34)     &T + P    &0.0708  &-0.766
   &$1.19\times 10^{-5}$\\
$\pi^+\pi^-$& ~~~~~~~" &T + P'   &0.      &-0.777	&$1.18\times 10^{-5}$\\
           & ~~~~~~~"     &T      &0.0     &-0.951    &$1.43\times 10^{-5}$ \\
\hline
           & (-0.28,0.24)     &T + P    &0.0573  &-0.967
   &$1.03\times 10^{-5}$\\
$\pi^+\pi^-$& ~~~~~~~" &T + P'   &0.0     &-0.97     &$1.03\times 10^{-5}$ \\
           & ~~~~~~~"     &T        &0.0     &-0.869    &$1.50\times 10^{-5}$
\\
\hline \hline
\end{tabular}

\eject

\end{document}